\documentclass[twocolumn,aps,prd,groupedaddress,superscriptaddress,amsmath,floatfix,amssymb,showpacs,noeprint,english]{revtex4-2}
\usepackage{graphicx}% Include figure files
\usepackage{dcolumn}% Align table columns on decimal point
\usepackage{bm}% bold math
\usepackage[T1]{fontenc}
\usepackage{mathrsfs}
\usepackage{amsbsy}
\usepackage{amstext}
\usepackage{amsmath, amsthm, amsfonts, amssymb, bbold}
\usepackage{setspace}
\usepackage{esint}  	
\usepackage{xcolor,colortbl}
\usepackage{float}
\usepackage[colorlinks,citecolor=blue,urlcolor=blue,linkcolor=blue,hypertexnames=true]{hyperref} 
\usepackage{babel}
\usepackage{booktabs}
\usepackage{tikz}

\usepackage{tikz}
\usepackage{quantikz}

\begin{document}

\title{Preparing a Thermofield Double State with Feedback Quantum Algorithms}

\author{Guilherme E. L. Pexe}
\email{guilherme.pexe@ifsc.usp.br}
\affiliation{Instituto de F\'{\i}sica de S\~ao Carlos, Universidade de S\~ao Paulo, CP 369, 13560-970 S\~ao Carlos, Brazil}

\author{Lucas A. M. Rattighieri}
\affiliation{Instituto de F\'{\i}sica Gleb Wataghin, Universidade Estadual de Campinas, 13083-859 Campinas, SP, Brazil}

\author{Felipe F. Fanchini}
\affiliation{Faculty of Sciences, UNESP - S{\~a}o Paulo State University, 17033-360 Bauru-SP, Brazil}
\affiliation{QuaTI - Quantum Technology \& Information, 13560-161 São Carlos-SP, Brazil}

\author{Dario Rosa}
\affiliation{ ICTP South American Institute for Fundamental Research
Instituto de Física Teórica, UNESP - Univ. Estadual Paulista
Rua Dr. Bento Teobaldo Ferraz 271, 01140-070, S{\~a}o Paulo, SP, Brazil}

\author{Amilson R. Fritsch}
\affiliation{Instituto de F\'{\i}sica de S\~ao Carlos, Universidade de S\~ao Paulo, CP 369, 13560-970 S\~ao Carlos, Brazil}

\date{\today}

\begin{abstract}
The efficient preparation of correlated thermal states, such as the Thermofield Double (TFD) state, is a fundamental prerequisite for simulating quantum gravity models and many-body thermodynamics on quantum processors. In this work, we investigate the ground state preparation of the Two Coupled Sachdev-Ye-Kitaev model, known as the Maldacena-Qi model, which is dual to a traversable wormhole in $AdS_2$, utilizing feedback-based quantum algorithms. We demonstrate that the standard feedback-based quantum algorithm (FALQON) and its time-rescaled variant (TR-FALQON) face severe kinetic limitations in this system, failing to converge to the highly entangled ground state when initialized in trivial product states. To overcome these barriers, we propose the hybrid ITE-TR-FALQON protocol, which integrates the imaginary-time evolution present in imaginary-time-enhanced FALQON (ITE-FALQON) with the time-rescaling mechanism. Our numerical results indicate that the introduction of non-unitary dynamics is strictly necessary to break symmetry traps and filter out excited states, while time-rescaling drastically accelerates algorithm convergence. The proposed method achieves fidelities close to unity and reproduces the von Neumann and Rényi entropy spectra of the exact TFD state with high precision.
\end{abstract}

\maketitle

\section{INTRODUCTION}

The Sachdev–Ye–Kitaev (SYK) model has garnered significant interest within the theoretical physics community due to its rich connections to quantum chaos, quantum gravity, and non-quasiparticle states of matter. In its original incarnation, it consists of a system of \(N\) Majorana fermions with all-to-all random \(q\)-body interactions sampled from a Gaussian distribution, exhibiting maximally chaotic dynamics \cite{PhysRevLett.70.3339, PhysRevD.94.106002}. In the low-energy regime, it admits a holographic dual description as Jackiw–Teitelboim gravity in \(AdS_2\) \cite{Sachdev_2015, PolchinskiRosenhaus2016Spectrum}. In the large-\(N\) limit, the SYK spectrum becomes effectively continuous, and the model saturates the Maldacena–Shenker–Stanford bound on Lyapunov exponents \cite{MaldacenaStanfordYang2016NearlyAdS2, KitaevSuh2018SoftMode}, serving as a theoretical laboratory for investigating quantum thermalization, entanglement entropy, and non-perturbative phenomena in nearly conformal field theories \cite{GuLucasQi2017Entanglement, Rosenhaus2019IntroSYK}. Furthermore, as an exactly solvable example of a strange metal, the SYK model has deepened our understanding of non-Fermi liquids and novel strongly correlated electronic phases \cite{Maldacena_2016, ChowdhuryGeorgesParcolletSachdev2022Review}.

Over the years, various extensions and generalizations of the model have been proposed. Among them, a prominent case is the Two Coupled SYK model, introduced by Maldacena and Qi \cite{maldacena2018eternaltraversablewormhole}, in which two identical copies of the SYK model (that are often referred to as ``Left'' and ``Right'', respectively) are connected by a coupling term, bilinear in the Majorana fermions and having strength parametrized by an extra parameter \(\mu\). For a certain range of $\mu$ values, this system exhibits a ground state that approximates the Thermofield Double (TFD) state between the left (L) and right (R) copies with high precision, with an inverse temperature $\beta$ functionally determined by the coupling strength. The bilinear interaction is given by (we denote the $N$ Majorana fermions belonging to the left system as $\chi_j^L$, and the Majorana fermions belonging to the right system as $\chi_j^R$)
\begin{equation}
  i \mu \sum_{j=1}^{N} \chi_j^L \chi_j^R.
\end{equation}
When the value of $\mu$ is properly tuned, it induces a bilateral correlation that, in the holographic interpretation, corresponds to two \(AdS_2\) geometries connected by a traversable wormhole \cite{maldacena2018eternaltraversablewormhole,PhysRevD.100.026002}. This arrangement has been explored as a prototype for studying the traversability of quantum wormholes and the emergence of gravity from strongly correlated quantum systems.

Beyond the profound interest in the context of quantum gravity, the construction and manipulation of TFD states is of paramount importance for the development of quantum information and emerging quantum technologies. The TFD state is a valuable resource for quantum teleportation protocols, serves as a basis for quantum error correction codes, and offers a robust benchmark for testing scrambling dynamics and information propagation in quantum processors \cite{Preskill_2018, susskind2014computationalcomplexityblackhole}. The ability to efficiently prepare such states is, therefore, a crucial step toward validating scalable quantum computing architectures.

However, the experimental preparation of states such as the TFD on quantum devices remains challenging. Conventional hybrid variational methods, such as the Variational Quantum Eigensolver (VQE), rely on the classical optimization of parameters in a parameterized circuit. This approach becomes computationally expensive as the number of qubits increases and faces severe obstacles, such as the Barren Plateau problem, where the gradient of the cost function vanishes exponentially with system size, rendering optimization unfeasible \cite{Liu_2024}. Moreover, achieving a faithful TFD state requires deep circuits, which are highly susceptible to noise in Noisy Intermediate-Scale Quantum (NISQ) hardware.

In this scenario, feedback-based quantum algorithms (FQA), and more specifically the FALQON (Feedback-based ALgorithm for Quantum OptimizatioN) protocol \cite{PhysRevA.106.062414, PhysRevB.110.224422}, emerge as a promising alternative. From the viewpoint of quantum gravity, this type of protocol is particularly appealing, since the efficient preparation of TFD states is directly connected to the study of traversable wormholes and to the holographic implementation of correlations between two coupled systems. FALQON is a closed-loop control approach that avoids classical variational optimization, updating circuit parameters deterministically from observables measured during the evolution. It therefore offers a promising route for preparing strongly entangled states on NISQ quantum devices. In these algorithms, circuit parameters are defined deterministically through rules inspired by Lyapunov Quantum Control, eliminating the need for an external classical optimizer and therefore mitigating the Barren Plateau problem. Although standard FALQON and its time-rescaled variant, the Time-Rescaled Feedback Quantum Algorithm (TR-FALQON), are efficient for many Hamiltonians \cite{qc91-5mj2}, previous work has already shown that standard FALQON can be inefficient for certain models, as in the case of the ANNNI model \cite{PhysRevB.110.224422}. In this work, we will show that standard FALQON is inefficient also for the Maldacena-Qi model when the circuit is initialized in trivial quantum computational states, such as product states with no prior correlations. This choice of initial state is natural, since quantum processors typically provide trivial initial states; therefore, to ensure a fair comparison, the algorithm should start from such states.

To overcome this limitation and enhance the method's convergence, we propose and develop in this work the hybrid ITE-TR-FALQON protocol. This new variant integrates imaginary-time evolution (ITE) \cite{vanlong2025imaginarytimeenhancedfeedbackbasedquantumalgorithms}, which introduces an effective non-unitary dynamics capable of filtering out excited states, with the time-rescaling (TR) mechanism, which dynamically adjusts the magnitude of evolution steps. The combination of these strategies allows the algorithm to escape from the kinetic traps and converge monotonically to the ground state of the coupled system, even when starting from a generic and trivial initial state where standard versions fail.

In this work, we implement the FALQON, TR-FALQON, ITE-FALQON variants, and the proposed combined ITE-TR-FALQON on the Maldacena-Qi model to prepare its entangled thermal state. Our main results demonstrate that: (i) the combined ITE-TR-FALQON approach is strictly necessary and superior for obtaining the ground state from uncorrelated initial states, overcoming the limitations of purely unitary versions and improving upon the efficiency of the standard ITE version; (ii) the prepared state exhibits high fidelity with the exact TFD and correctly reproduces entanglement properties, verified through the calculation of von Neumann and Rényi entropies. We also analyze the robustness of the algorithms against variations in control parameters, establishing a viable path for the preparation of complex thermodynamic states on quantum hardware.

This paper is organized as follows. In Sec.~\ref{FQA}, we review the feedback-based quantum algorithms employed in this work, including the TR-FALQON and ITE-FALQON variants, and introduce the hybrid ITE-TR-FALQON protocol proposed here. In Sec.~\ref{Maldacena-Qi}, we review the Maldacena-Qi model and define the quantities used to characterize the quality of the prepared state, in particular the fidelity and the entanglement entropies. In Sec.~\ref{results}, we present the numerical results for the preparation of the ground state of the model and compare the performance of the different algorithms. Finally, in Sec.~\ref{Conclusion}, we summarize our conclusions and discuss possible directions for future work.

\section{Feedback Quantum Algorithms} \label{FQA}

\begin{figure*}[htpb]
    \centering
    \includegraphics[width=1\linewidth]{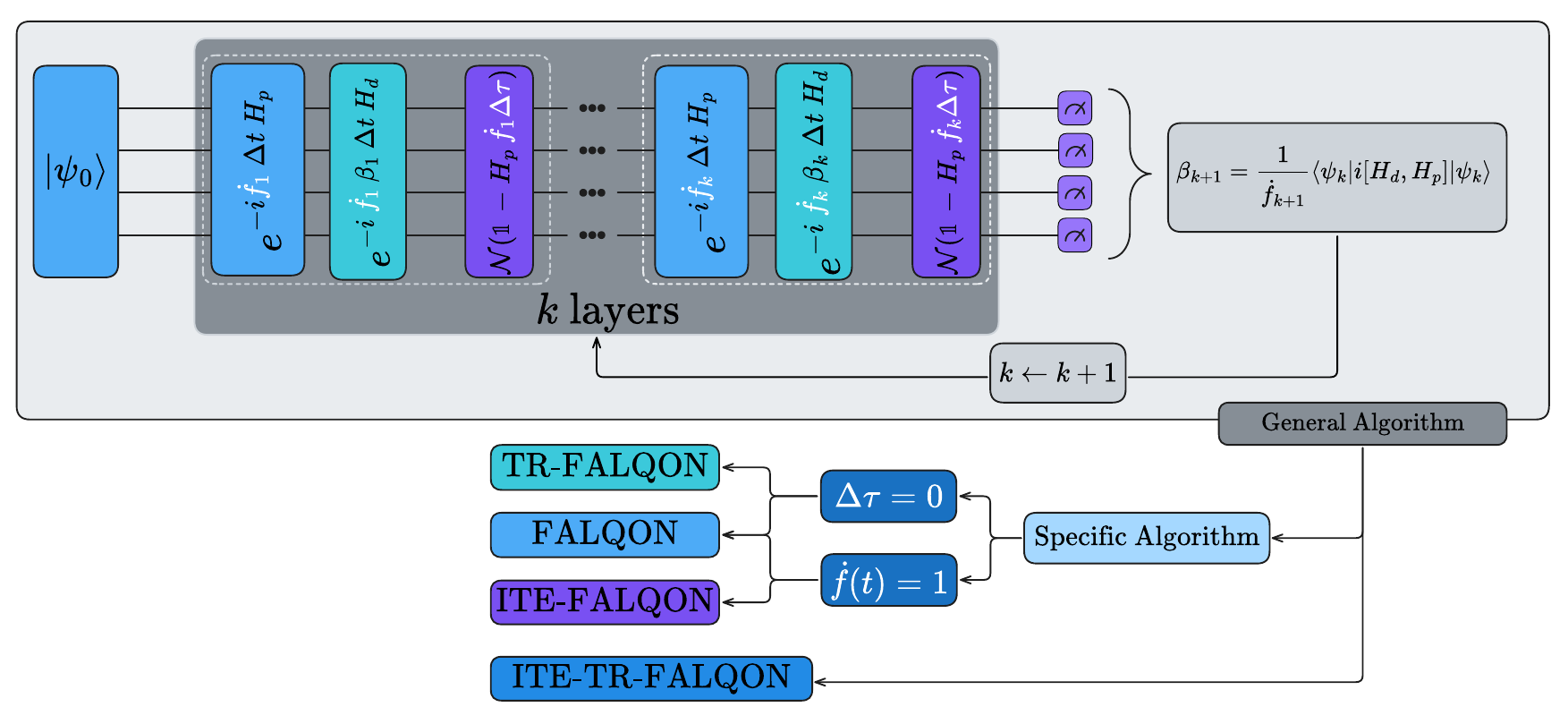}
    \caption{Unified schematic representation of the FALQON protocols. The upper diagram illustrates the General Algorithm, where the initial state $|\psi_0\rangle$ evolves through $k$ layers. Each layer sequentially applies unitary operators generated by $H_p$ and $H_d$, modulated by the time-rescaling factor $\dot f_k$ and the time step $\Delta t$, followed by a non-unitary operator $\mathcal{N}(1 - H_p\Delta\tau)$ associated with imaginary-time evolution. The parameter $\beta_{k+1}$ is calculated deterministically at each step via the expectation value of the commutator $\langle i[H_d,H_p]\rangle$. The lower flowchart demonstrates how the General Algorithm specializes into specific methods: the condition $\Delta\tau=0$ removes the non-unitary component, recovering FALQON and TR-FALQON (the latter dependent on the modulation of $\dot f$); alternatively, fixing $\dot f(t)=1$ preserves the dissipative dynamics, characterizing the ITE-FALQON protocol.}
    \label{fig:diag_falqon}
\end{figure*}

The FALQON algorithm relies on Lyapunov Quantum Control (QLC)~\cite{PhysRevA.106.062414} to design controls that asymptotically guide a quantum system toward a target state. In this context, we consider the time-dependent Hamiltonian
\begin{equation}
    H(t) = H_p + \beta(t) H_d,
\end{equation}
where $H_p$ is the problem Hamiltonian, namely the Hamiltonian whose ground state we aim to prepare, and $H_d$ is the driver Hamiltonian, which induces the dynamics during the evolution. These names follow the standard terminology used in the literature, as in~\cite{1Magann_2022}. The corresponding Schrödinger equation (with $\hbar = 1$) is

\begin{equation}
    i \frac{d}{dt} \ket{\psi(t)} = (H_p + \beta(t) H_d) \ket{\psi(t)}.
    \label{eqn:equacao_sch}
\end{equation}
Defining the Lyapunov function $J(t)=\bra{\psi(t)}H_p\ket{\psi(t)}$, one aims to design $\beta(t)$ to ensure
\begin{equation}
    \frac{dJ(t)}{dt} \le 0, \quad \forall t,
    \label{eqn:condicao_lyapunov}
\end{equation}
which in turn guarantees the approach to the ground state.
Calculating the derivative yields $\frac{dJ}{dt}=A(t)\beta(t)$, with $A(t)=\bra{\psi(t)} i[H_d,H_p]\ket{\psi(t)}$. Therefore, a choice that guarantees the validity of Eq.~\eqref{eqn:condicao_lyapunov} is
\begin{equation}
    \beta(t) = -A(t),
    \label{eqn:beta_especifico}
\end{equation}
provided that $[H_d,H_p]\neq 0$. Instead of relying on an external classical optimizer, FALQON iteratively constructs a Trotterized circuit and updates, via feedback, the parameters of each layer based on measurements of the state generated by previous layers. The formal solution discretized by Suzuki–Trotter in $k$ steps ($\Delta t$) is
\begin{equation}
\ket{\psi_k} = U_d(\beta_k) U_p \dots U_d(\beta_1) U_p \ket{\psi_0},
\label{eq_ev_trottenizada}
\end{equation}
with $U_d(\beta_k)=e^{-i\beta_k H_d \Delta t}$ and $U_p=e^{-iH_p\Delta t}$. The feedback law used in the layer construction is
\begin{equation}
    \beta_k = - A_{k-1} = - \bra{\psi_{k-1}} i[H_d, H_p] \ket{\psi_{k-1}},
    \label{eqn:lei_feedback0}
\end{equation}
where $A_{k-1}$ is estimated from the state of the previous iteration. The procedure is repeated until $J_k=\langle\psi_k|H_p|\psi_k\rangle$ converges. Figure \ref{fig:diag_falqon} illustrates the diagram of the method and its variants.

\subsection{TR-FALQON} \label{TR-FALQON}
The TR-FALQON protocol \cite{qc91-5mj2, PhysRevResearch.2.013133, ferreira2024shortcutsadiabaticitydesignedtimerescaling} modifies the standard FALQON algorithm just described by applying an extra time variable change $t=f(t)$, where the evolution satisfies
\begin{equation}
    \frac{d}{dt}\ket{\psi(t)} = -i H(f(t))\dot f(t)\ket{\psi(t)} \equiv -i \tilde{H}(t)\ket{\psi(t)}.
\end{equation}
Repeating the calculation of $\frac{d}{dt}\langle H_p\rangle$ yields a term proportional to $\tilde{\beta}(t)\dot f(t) A(t)$, with $A(t)=\langle\psi(t)|i[H_d,H_p]|\psi(t)\rangle$ and $\tilde{\beta}(t)=\beta(f(t))$. Choosing appropriate continuous functions, the control law becomes
\begin{equation}
    \tilde{\beta}(t) = - A(t)\frac{1}{\dot f(t)},
\end{equation}

which preserves the monotonicity of $J(t)$. Discretizing into steps $\Delta \tau$ and Trotterizing, the rescaled evolution is given by
\begin{equation}
    \tilde{U}(t)=\tilde{U}_d(\tilde{\beta}(t))\,\tilde{U}_p,\quad 
    \tilde{U}_d(\tilde{\beta})=\exp\!\big(-iH_d\,\tilde{\beta}(\tau)\dot{f}(\tau)\Delta \tau\big),
\end{equation}
with
\begin{equation}
    \tilde{U}_p=\exp\!\big(-iH_p\,\dot{f}(\tau)\Delta \tau\big),
\end{equation}

leading to the circuit

\begin{equation}
    \ket{\psi_k} = \tilde{U}_d(\tilde{\beta}_k)\tilde{U}_p\;\dots\;\tilde{U}_d(\tilde{\beta}_1)\tilde{U}_p\ket{\psi_0},
\end{equation}
with the discretized feedback law
\begin{equation}
    \tilde{\beta}_k = - A_{k-1}\,\frac{1}{\dot{f}(k\Delta t)}.
    \label{eqn:lei_feedback}
\end{equation}
TR-FALQON uses this transformation to accelerate convergence and reduce circuit depth, with the choice of $f(t)$ influencing performance depending on the problem.

A common choice for the rescaling function is:
\begin{equation}
f(t) = at - \frac{t_f}{2\pi a}(a-1)\sin\left(\frac{2\pi at}{t_f}\right)
\end{equation}
where $a$ is a parameter controlling time contraction. Its inverse function $f^{-1}(t)$, which cannot be expressed exactly in terms of standard functions, satisfies the properties $f^{-1}(0) = 0$, $f^{-1}(t_f) = \frac{t_f}{a}$, $f'(0) = 1$, and $f'(\frac{t_f}{a}) = 1$, which can be easily verified. These elements qualify this function as an appropriate time-rescaling function for any $a > 1$ \cite{qc91-5mj2}.

\subsection{ITE-FALQON} \label{ITE-FALQON}
The ITE-FALQON protocol, proposed by Van Long et al. \cite{vanlong2025imaginarytimeenhancedfeedbackbasedquantumalgorithms}, introduces an effective dissipative dynamics that suppresses excited state amplitudes \cite{ite1, ite2, ite3, ite4, ite5, ite6}. The imaginary-time Schrödinger equation,
\begin{equation}
\partial_{\tau}\ket{\psi(\tau)} = -H_p\ket{\psi(\tau)},
\end{equation}
possesses a formal solution analogous to diffusion:
\begin{equation}\label{eq:ite_solution}
\ket{\psi(\tau)} = e^{-H_p\tau}\ket{\psi(0)} = \sum_n c_n e^{-E_n\tau}\ket{n}.
\end{equation}
Higher energy components decay exponentially faster, causing $\ket{\psi(\tau)}$ to converge to the ground state of $H_p$ as $\tau\to\infty$. Applying the Taylor expansion for finite steps, one obtains
\begin{equation}\label{eq:taylor}
\ket{\Phi_{\Delta\tau}(\psi)}\approx \frac{(I-H_p\Delta\tau)\ket{\psi(\tau)}}{\|(I-H_p\Delta\tau)\ket{\psi(\tau)}\|}.
\end{equation}
ITE-FALQON inserts the operation from Eq.~\eqref{eq:taylor} periodically between unitary steps, resulting in
\begin{equation}\label{eq:psi_k}
\ket{\psi_k} = \mathcal{N}\bigl[(I-H_p\Delta\tau)\,e^{-iH_p\Delta t}\,e^{-i\beta_{k-1}H_d\Delta t}\,\ket{\psi_{k-1}}\bigr],
\end{equation}
where $\mathcal{N}$ is the normalization factor. The term $(I-\Delta\tau H_p)$ redistributes the population to lower energy sectors, reactivating the feedback term. Let $H_p$ be bounded such that $\|H_p\| \le h$. The hybrid algorithm interleaves ITE updates with the feedback control $\beta(t) = -i \langle [H_d, H_p] \rangle_t$. It is important to note that, within the scope of this work, we evaluate the algorithmic necessity of this non-unitary step relying on exact state-vector simulations. The concrete implementation of Eq.~\eqref{eq:taylor} on actual quantum hardware, which fundamentally requires embedding the non-unitary operator into a larger unitary space via ancilla qubits and post-selection, is highly non-trivial and is discussed further in Sec.~\ref{Conclusion}. At each ITE update, the normalized cost function is given by
\begin{equation}
C\bigl(\Phi_{\Delta\tau}(\psi)\bigr)
=
\frac{
\langle \psi | (I - \Delta\tau H_p)\, H_p \, (I - \Delta\tau H_p) | \psi \rangle
}{
\langle \psi | (I - \Delta\tau H_p)^2 | \psi \rangle
} .
\label{eq:cost_update_correct}
\end{equation}
Furthermore, one has the upper bound
\begin{equation}
C\bigl(\Phi_{\Delta\tau}(\psi)\bigr)
\le
E
-
\frac{
2 \Delta\tau \left( 1 - \tfrac{1}{2} h \Delta\tau \right)
}{
(1 - h \Delta\tau)^2
}
\, \mathcal{V} ,
\label{eq:bound_correct}
\end{equation}
where $E = \langle H_p \rangle_\psi$ and $\mathcal{V} = \langle (H_p - E)^2 \rangle_\psi$. The bound holds for $0 < \Delta\tau < 2/h$ (with $\Delta\tau \neq 1/h$). If $\mathcal{V} > 0$, then $C\bigl(\Phi_{\Delta\tau}(\psi)\bigr) < C(\psi)$, ensuring that the sequence of function values strictly decreases until reaching the ground state ($\mathcal{V} = 0$) \cite{vanlong2025imaginarytimeenhancedfeedbackbasedquantumalgorithms}.

\section{Maldacena-Qi model} \label{Maldacena-Qi}

The Maldacena-Qi model consists of two identical copies of the Sachdev-Ye-Kitaev model, denoted by $L$ (Left) and $R$ (Right), coupled through a bilinear interaction term. For Majorana fermions, the total Hamiltonian is expressed as:
\begin{equation}
H = H_L + H_R + i\,\mu \sum_{j=1}^N \chi^L_j \chi^R_j,
\label{eq:H_total}
\end{equation}
where $N$ is the number of Majorana fermions in each copy and $\mu$ represents the coupling strength. The individual Hamiltonians for each copy, for an interaction order $q$, are given by:
\begin{equation}
H_{\alpha} = i^{q/2} \sum_{1\le j_1<\dots<j_q\le N} J_{j_1\dots j_q}^{\alpha}\,\chi^\alpha_{j_1} \dots \chi^\alpha_{j_q}, \quad (\alpha = L, R).
\end{equation}
The Majorana operators satisfy the Clifford algebra $\{\chi^\alpha_j,\chi^\beta_k\} = 2\delta^{\alpha\beta}\delta_{jk}$. In this work, we focus on the $q=4$ case. As numerically implemented, the symmetry between the copies correlates the random couplings such that $J_{j_1\dots j_q}^{R} = (-1)^{q/2} J_{j_1\dots j_q}^{L}$. For $q=4$, this results in identical couplings for both copies ($J_{j_1\dots j_q}^L = J_{j_1\dots j_q}^R \equiv J_{j_1\dots j_q}$).

The coefficients $J_{j_1\dots j_q}$ are independent random variables drawn from a Gaussian distribution with zero mean and variance defined by:
\begin{equation}
\overline{J_{j_1\dots j_q}^2} = \frac{2^{q-1}(q-1)!}{q N^{q-1}} \mathcal{J}^2,
\label{eq:variancia}
\end{equation}
where $\mathcal{J}$ sets the microscopic energy scale, fixed as $\mathcal{J}=1$ in our simulations.

In the weak but finite coupling regime, the ground state of this Hamiltonian, $|\psi_0\rangle$, approximates with high fidelity a Thermofield Double (TFD) state \cite{maldacena2018eternaltraversablewormhole}:
\begin{equation}
|\mathrm{TFD}(\beta)\rangle = \frac{1}{\sqrt{Z(\beta)}} \sum_n e^{-\beta E_n/2} \,|n\rangle_L \otimes |\bar{n}\rangle_R,
\end{equation}
where $|n\rangle$ are the energy eigenstates of the isolated SYK system and $\beta$ is the effective inverse temperature determined by the coupling $\mu$.

\subsection{Fidelity} \label{Fidelity}

To quantify the effectiveness of the feedback-based quantum algorithm (FALQON) in preparing the ground state, we calculate the fidelity between the state obtained via the algorithm, $|\psi_{\rm FALQON}\rangle$, and the exact ground eigenvector obtained by numerical diagonalization, $|\psi_{0}\rangle$:
\begin{equation}
F = |\langle \psi_{\rm FALQON} | \psi_{0} \rangle|^2.
\label{eq:fidelidade}
\end{equation}
Values of $F \approx 1$ confirm the convergence of the algorithm to the vacuum of the coupled system.

\subsection{Entanglement Entropy} \label{Entanglement Entropy}

The characterization of quantum entanglement is fundamental to validating the holographic and thermodynamic nature of the prepared state. In the context of the AdS/CFT duality and the coupled SYK model, the entanglement between copies $L$ and $R$ in the TFD state is not merely a measure of non-local correlations, but the geometric manifestation of the connectivity of the dual spacetime (the Einstein-Rosen bridge).

To quantify these correlations, we perform a bipartition of the total system, defining the reduced density matrix of one of the subsystems (e.g., the $L$ side) as $\rho_L = \operatorname{Tr}_{R}(|\psi\rangle\langle\psi|)$. The primary measure of interest is the von Neumann entropy:
\begin{equation}
S^{(\mathrm{vN})}_L = -\operatorname{Tr}(\rho_L \ln \rho_L).
\end{equation}
Physically, for an ideal TFD state, $S^{(\mathrm{vN})}_L$ corresponds to the thermodynamic entropy of the SYK system at inverse temperature $\beta$. In the holographic context, this quantity is dual to the area of the minimal surface in the bulk (Ryu-Takayanagi formula), serving as a direct diagnostic of the geometry of the formed wormhole.

Complementarily, we calculate the Rényi entropies of order $n$:
\begin{equation}
S^{(n)}_L = \frac{1}{1-n}\ln \operatorname{Tr}(\rho_L^n).
\end{equation}
The inclusion of Rényi entropies is strategically crucial in this work for two main reasons.
First, from the perspective of quantum computing implementation, the von Neumann entropy is computationally expensive to estimate experimentally, requiring full state tomography. In contrast, the second-order Rényi entropy ($S^{(2)}_L$), related to subsystem purity, is directly accessible on quantum processors (NISQ) via efficient protocols such as the Swap Test or randomized measurements.

Second, the family of entropies $S^{(n)}_L$ probes the fine structure of the entanglement spectrum (the eigenvalues of $\rho_L$). For the TFD state, this spectrum must follow a Boltzmann thermal distribution. By verifying the agreement of Rényi entropies for different orders $n$, we confirm that the algorithm not only reproduces the correct average entanglement but faithfully captures the thermal statistics of the state, which is a prerequisite for gravitational duality.

\section{Results} \label{results}

In this section, we demonstrate how a quantum computer can be employed to investigate strongly interacting and long-range chaotic fermionic systems using the FALQON algorithm and its variants. The section is divided into subsections. First, we show how to employ FALQON to calculate the ground state. Next, we illustrate how this tool can be used to investigate state fidelity and entanglement entropy.

To implement the problem Hamiltonian $H_p$, we map the fermionic degrees of freedom onto a qubit register using the Jordan-Wigner transformation \cite{Bravyi_2002, PhysRevLett.119.040501}. In the notation adopted in this work, $N$ denotes the number of Majorana fermions in each SYK copy, so that the full Maldacena-Qi system contains $2N$ Majorana modes in total. After the Jordan-Wigner mapping, the model is encoded into a register of $N_q=N$ qubits. In the convention used in this work, the Majorana operators are written as
\begin{equation}
\begin{aligned}
\chi_{2j-1} &=
\left(\prod_{k=1}^{j-1}\sigma_k^z\right)\sigma_j^x, \\
\chi_{2j} &=
\left(\prod_{k=1}^{j-1}\sigma_k^z\right)\sigma_j^y,
\qquad j=1,\dots,N_q.
\end{aligned}
\end{equation}
Equivalently, the ordering of the Majorana modes in Eq.~(\ref{eq:H_total}) is identified with this standard Jordan-Wigner convention. Under this transformation, each Majorana operator is represented by a nonlocal Pauli string that preserves the original Clifford algebra. Consequently, both the four-body interaction terms and the bilinear coupling defined in Eq.~(\ref{eq:H_total}) are converted into a sum of tensor products of Pauli operators $\{I, X, Y, Z\}^{\otimes N_q}$, allowing for the explicit construction of the Hamiltonian matrix for simulating time evolution and evaluating the cost function.

To employ FALQON, TR-FALQON, and ITE-FALQON as tools for preparing the Thermofield Double (TFD) type ground state of the Maldacena-Qi model, we define the driver Hamiltonian $H_d$ as
\begin{equation}
H_d = \sum_{j=1}^{N_q} \sigma_j^x.
\end{equation}
Here, $\sigma^x_j$ is the Pauli-X matrix acting on qubit $j$. In the computational basis $\{\ket{0},\ket{1}\}$, the $\sigma^x$ matrix has the representation
\begin{equation}
\sigma^x = \begin{pmatrix} 0 & 1 \\[2pt] 1 & 0 \end{pmatrix},
\end{equation}
such that
\begin{equation}
\sigma^x \ket{0} = \ket{1}, \qquad \sigma^x \ket{1} = \ket{0}.
\end{equation}

The choice of the initial state used in our experiments is the product state
\begin{equation}
\ket{\psi_0} = \bigotimes_{j=1}^{N_q} \ket{+}_j \equiv \ket{+}^{\otimes N_q},
\end{equation}
that is, each qubit is prepared in the $\ket{+}$ state. We justify this choice for the following practical and conceptual reasons: (i) $\ket{+}^{\otimes N_q}$ is easy to prepare on quantum hardware (by applying Hadamard gates to all qubits starting from $\ket{0}^{\otimes N_q}$); (ii) it is a uniform superposition over the computational basis, thus introducing no initial bias towards specific classical configurations and allowing the algorithm to broadly explore the Hilbert space; (iii) being a product state with no initial entanglement, any entanglement generated during evolution is solely the result of interactions and the preparation protocol, which facilitates the interpretation of entanglement entropy measurements.

Finally, we emphasize that although $H_d=\sum_j \sigma_j^x$ is the standard mixer and offers implementation advantages, there are alternatives that may be preferable in scenarios with conservation constraints or reduced state spaces. We opted for the above definition for consistency with the literature and to facilitate direct comparisons between our results and related works.

\subsection{Ground State Preparation and Fidelity}

To evaluate the efficiency of the proposed protocols in preparing the ground state of the Maldacena-Qi model, we analyze the energy convergence and the fidelity of the prepared state as a function of the number of layers $k$. Simulations were performed for a system with $N=8$ Majorana fermions per SYK copy ($2N=16$ Majorana fermions in total, corresponding to $N_q=8$ qubits after the Jordan-Wigner mapping), considering different coupling regimes $\mu$.

\begin{figure}[!htbp]
    \centering
    \includegraphics[width=1\linewidth]{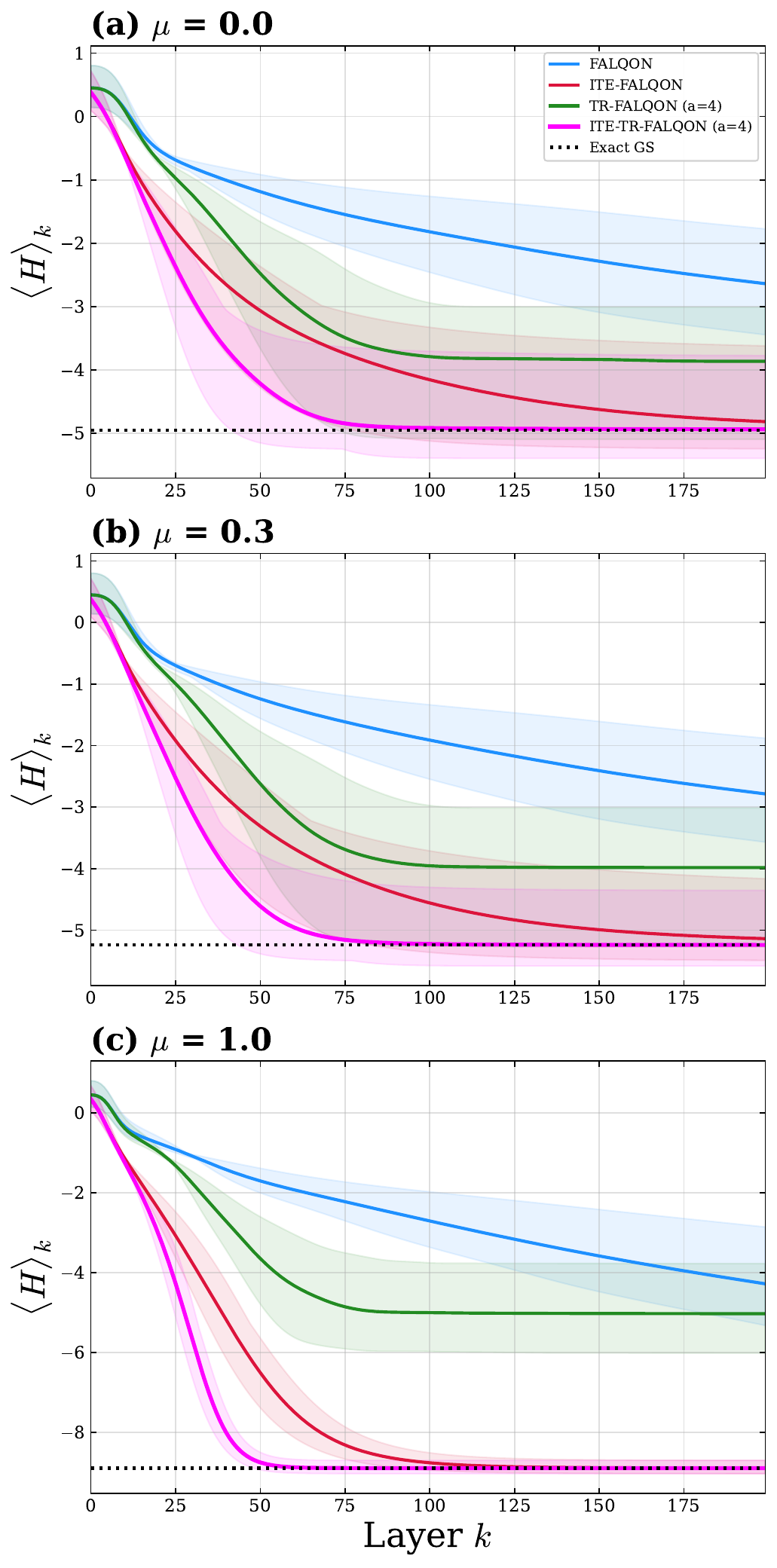}
    \caption{Convergence of the expected energy value $\langle H \rangle_k$ as a function of the number of layers $k$ for $N=8$ Majorana fermions per SYK copy. Panels show different coupling strengths: (a) $\mu=0.0$, (b) $\mu=0.3$, and (c) $\mu=1.0$. Curves represent the average of 10 realizations, with shaded areas indicating the standard deviation. The black dotted line indicates the exact ground state energy obtained by diagonalization. Simulations use $\Delta t = 0.01$ and $\Delta\tau = 0.01$.}
    \label{fig:comparacao}
\end{figure}

Fig.~\ref{fig:comparacao} presents the evolution of the energy expectation value $\langle H \rangle_k$ across layers for three distinct coupling strengths. The curves represent the average over disorder realizations, comparing the standard FALQON, ITE-FALQON, TR-FALQON algorithms, and the hybrid ITE-TR-FALQON method.

It is observed that for the uncoupled case ($\mu=0.0$) and weak coupling ($\mu=0.3$), the purely unitary methods (FALQON and TR-FALQON with $\alpha=4$) manage to reduce the energy but exhibit a slow convergence rate and a tendency to saturate above the ground state. TR-FALQON demonstrates a steeper initial descent than standard FALQON due to time-rescaling, which effectively amplifies gradients at the beginning of the evolution, yet it still fails to reach the exact energy within the simulated circuit depth.

The introduction of non-unitary dynamics (ITE) qualitatively alters the convergence behavior. As evidenced in panels (b) and (c) of Fig.~\ref{fig:comparacao}, as the coupling $\mu$ increases—making the optimization landscape more complex due to correlations between systems $L$ and $R$—the need to filter out excited states becomes critical. ITE-FALQON achieves significantly lower energies than the unitary versions. However, it is the hybrid ITE-TR-FALQON protocol that displays superior performance in all scenarios. By combining the exponential filtering of excited states from imaginary-time evolution with the kinetic acceleration provided by Time-Rescaling ($\alpha=4$), the algorithm reaches the exact ground state ($\langle H \rangle_k \approx E_{GS}$) in fewer layers for $\mu=1.0$, a regime where standard methods fail or stagnate in local minima.

\begin{figure}[!htbp]
    \centering
    \includegraphics[width=1\linewidth]{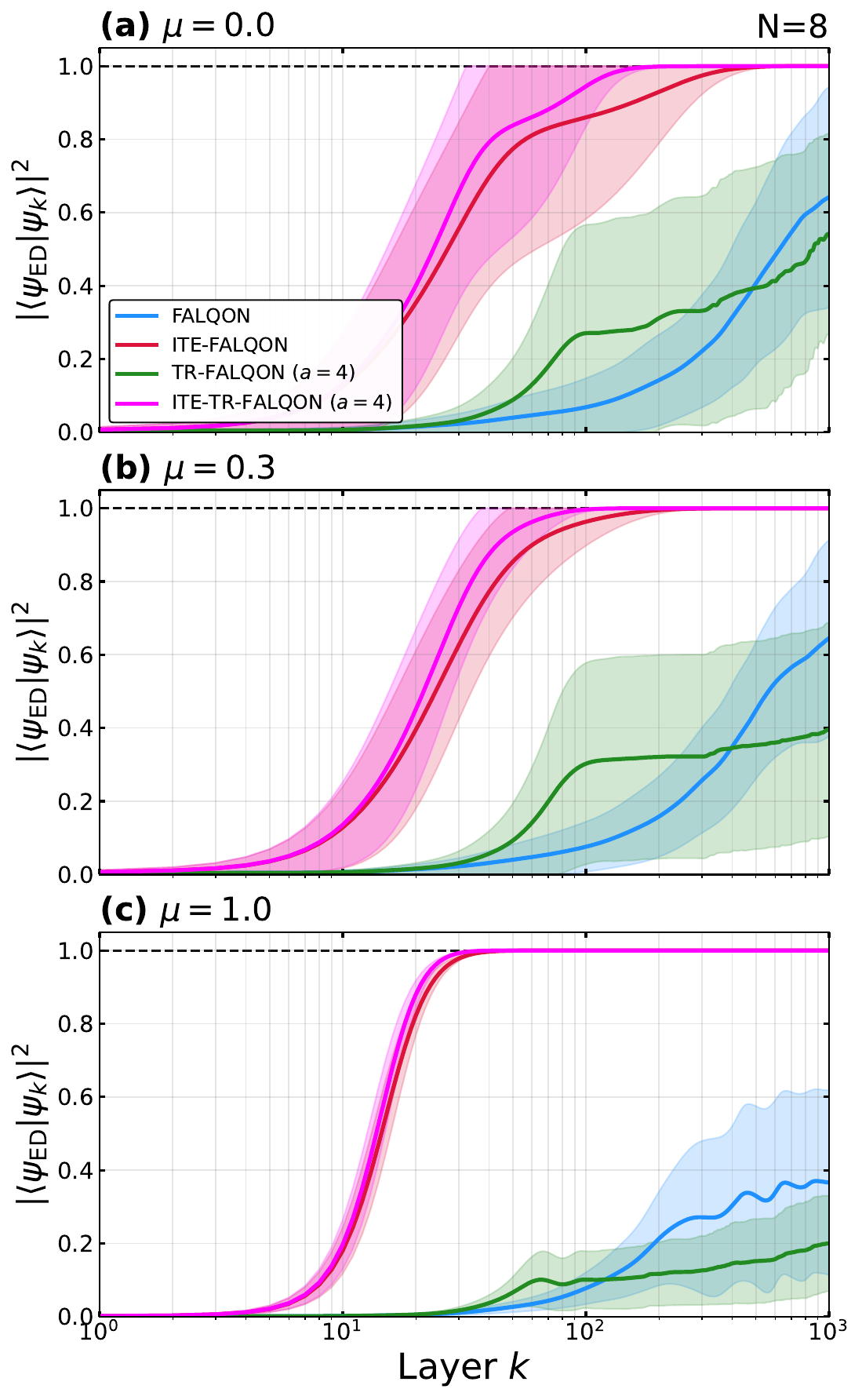}
    \caption{Fidelity of the prepared state $|\langle \psi_{\text{ED}} | \psi_k \rangle|^2$ relative to the exact ground state on a logarithmic scale for $N=8$ Majorana fermions per SYK copy. The panels show three different coupling strengths: (a) $\mu=0.0$, (b) $\mu=0.3$, and (c) $\mu=1.0$. Comparison is made between FALQON, ITE-FALQON, TR-FALQON, and ITE-TR-FALQON methods. The time-rescaling parameter is fixed at $a=4$. Curves represent the average of 10 realizations, with shaded areas indicating the standard deviation. Simulations use $\Delta t = 0.01$ and $\Delta\tau = 0.09$.}
    \label{fig:fidelity}
\end{figure}

To quantify the quality of the prepared quantum state, we calculate the fidelity $|\langle \psi_{\text{ED}} | \psi_k \rangle|^2$ relative to the ground state obtained by exact diagonalization. Fig.~\ref{fig:fidelity} illustrates the fidelity on a logarithmic scale of layers (up to $k=10^3$) for three different coupling strengths: $\mu=0.0$ (a), $\mu=0.3$ (b) and $\mu=1.0$ (c).

Analysis of Fig.~\ref{fig:fidelity} reveals a clear dichotomy between the methods across all observed regimes. Purely unitary algorithms (FALQON and TR-FALQON) remain trapped in low-fidelity states even after $10^3$ layers, indicating that unitary dynamics alone is incapable of efficiently escaping the subspace orthogonal to the target TFD state when starting from a trivial product state. In contrast, ITE-based methods exhibit a dynamical phase transition, where fidelity jumps abruptly to unity regardless of the coupling strength.

The impact of the time-rescaling parameter $a=4$ on the hybrid method is evident in all three panels: while ITE-FALQON requires approximately $10^2$ layers to initiate the transition to high fidelity, ITE-TR-FALQON anticipates this convergence, reaching fidelity $> 0.99$ significantly earlier. This result confirms that time-rescaling acts not only on energy minimization but also accelerates the rotation of the state vector toward the ground state subspace across different values of $\mu$, drastically reducing the circuit depth required to prepare high-quality TFD states.

\subsection{Entanglement Entropy}

The entanglement entropy between subsystems $L$ and $R$ offers a direct probe of the emergent geometry and the thermality of the prepared state. In the context of holographic duality, the von Neumann entropy $S_{\mathrm{vN}}$ of subsystem $L$ is dual to the area of the minimal surface in the bulk (Ryu-Takayanagi surface). To validate whether the state obtained by variational algorithms genuinely corresponds to the Thermofield Double state, we analyze both the entropy convergence and the full spectrum of Rényi moments.

\begin{figure}[!htbp]
    \centering
    \includegraphics[width=0.48\textwidth]{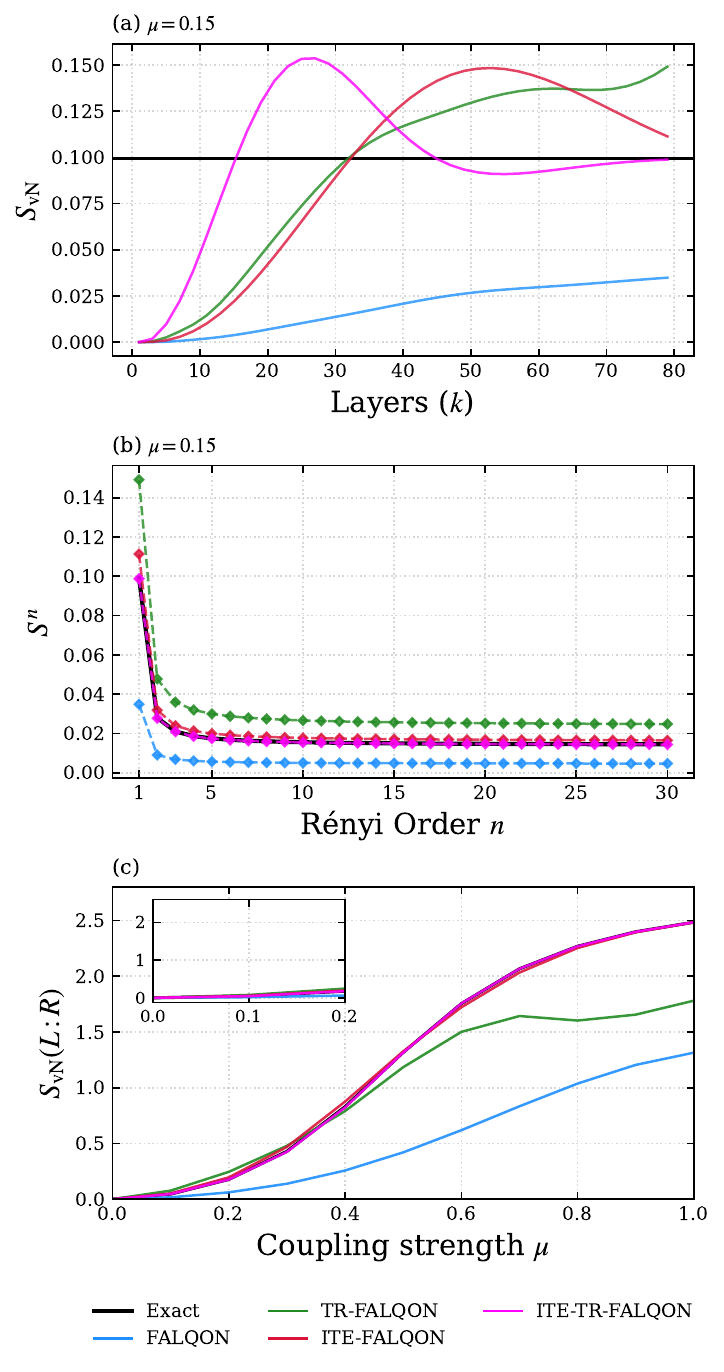} 
    \caption{\label{fig:entropia_analise} 
    Entanglement characterization of the ground state prepared for the coupled SYK model with $N=8$ Majorana fermions per SYK copy. 
    (a) Evolution of von Neumann Entropy $S_{\mathrm{vN}}(L:R)$ as a function of circuit layers ($k$) for $\mu=0.15$. The hybrid ITE-TR-FALQON method (magenta) reaches the exact value, while purely unitary methods saturate prematurely.
    (b) Spectrum of Rényi Entropies $S^n$ as a function of order $n$ for the final state obtained (state at $k=80$ for $\mu=0.15$). The coincidence with the exact curve (black) confirms the thermal distribution of the reduced density matrix eigenvalues.
    (c) Dependence of the steady-state von Neumann Entropy on the coupling strength $\mu$. The proposed algorithm correctly captures the transition from a product state ($\mu=0$) to a highly entangled state ($\mu=1$). 
    Parameters: time step $\Delta t = 0.02$, time-scaling factor $a=4$ (for TR variants), and ITE rate $\Delta\tau=0.2$.}
\end{figure}

Fig.~\ref{fig:entropia_analise}(a) presents the convergence dynamics of $S_{\mathrm{vN}}$ as a function of the number of layers ($k$) for a system with $N=8$ Majorana fermions per SYK copy and coupling $\mu=0.15$. A clear distinction is observed between purely unitary methods and those assisted by imaginary-time evolution (ITE). The FALQON and TR-FALQON algorithms quickly saturate at entropy values lower than the exact value, indicating that unitary evolution, restricted to conserved symmetry subspaces, struggles to access the full support of the TFD state in Hilbert space. In contrast, dissipative variants, specifically ITE-TR-FALQON, converge monotonically to the exact ground state entropy. This confirms that the non-unitary component is essential to "break" kinetic traps and allow the system to develop the necessary long-range correlations.

A more rigorous verification of the state's thermal nature is provided by the spectrum of Rényi Entropies, $S^n = (1-n)^{-1} \ln \mathrm{Tr}(\rho_L^n)$, shown in Fig.~\ref{fig:entropia_analise}(b). While $S_{\mathrm{vN}}$ provides a single scalar number, the dependence of $S^n$ on the order $n$ encodes the distribution of the reduced density matrix eigenvalues $\rho_L$. The perfect agreement between the ITE-TR-FALQON result and the exact curve for all orders $n$ demonstrates that the algorithm not only captures the total amount of entanglement but faithfully reproduces the Boltzmann statistics of the TFD state's Schmidt coefficients. The discrepancies observed for unitary methods at higher orders ($n > 5$) reveal that, although the energy may be close to the ground state, the fine structure of the entanglement differs significantly from the thermal target.

Finally, we investigate in Fig.~\ref{fig:entropia_analise}(c) the robustness of state preparation under variation of the coupling strength $\mu \in [0, 1.0]$. It is crucial to note that the physically relevant regime for the holographic interpretation of a traversable wormhole in $\mathrm{AdS}_2$ occurs at weak coupling. As established in \cite{PhysRevD.100.026002}, for large values of the coupling (e.g., $\mu \approx 1.0$), the direct one-body interaction between the two SYK models strongly dominates over the many-body term. In this regime, the system becomes essentially trivial and non-interacting, characterized by an energy gap that scales linearly with the coupling.

Furthermore, the detailed study of the energy gap and the Schwinger-Dyson equations performed in \cite{PhysRevD.100.026002} demonstrated that the non-trivial phase is located below a critical coupling $\mu_c \approx 0.175$. Below this threshold, the system genuinely dualizes to a traversable wormhole and the energy gap exhibits a fractional power-law behavior. Based entirely on the theoretical limits found in that reference, we selected $\mu = 0.15$ (well within the stable wormhole phase) as the strictly physically motivated benchmark for our detailed entanglement entropy calculations.

To properly highlight the performance of our approach in this region of actual holographic interest, we include an inset in Fig.~\ref{fig:entropia_analise}(c) magnifying the interval $\mu \in [0, 0.2]$. Our results show that the ITE-TR-FALQON protocol accurately tracks the exact theoretical curve precisely within this gap-stabilized regime, evidencing its applicability for preparing the relevant vacuum states of the model without being restricted to the trivial free-fermion limit.

\section{Conclusion} \label{Conclusion}

In this work, we addressed the challenge of preparing the ground state of the Maldacena-Qi model on quantum computers, a system of central interest for the study of AdS/CFT holographic duality and quantum chaos. We comparatively investigated the performance of different classes of Feedback-based Quantum Algorithms (FQA), focusing on the capacity of these methods to generate the Thermofield Double state from trivial, uncorrelated initial states.

Our analysis revealed a critical limitation in purely unitary methods. We demonstrated that standard FALQON and TR-FALQON, although efficient for various Hamiltonians, fail to prepare the ground state of the coupled SYK model in the strong coupling region ($\mu \approx 1.0$). We observed that unitary evolution tends to remain confined in subspaces orthogonal to the target state or saturate in local energy minima, incapable of generating the necessary long-range entanglement starting from a product state.

To circumvent this obstacle, we established that the introduction of effective non-unitary dynamics is essential. The implementation of the hybrid ITE-TR-FALQON algorithm proved to be the superior strategy. Our results show that this protocol not only resolves the convergence problem, reaching the exact ground state where standard versions fail, but does so with significantly higher efficiency than pure imaginary-time evolution (ITE), thanks to the kinetic acceleration provided by time-rescaling.

Beyond energetic validation, characterization via quantum information theory confirmed the physical quality of the prepared state. The ITE-TR-FALQON protocol faithfully reproduced the von Neumann entropy and the entire spectrum of Rényi entropies of the TFD state. This result is of paramount importance for holographic interpretation, as it confirms that the algorithm correctly captures the thermal statistics and the geometry of the dual wormhole, in contrast to unitary methods, which produce states with spurious entanglement signatures.

It is important to emphasize that while the current work establishes the algorithmic necessity of non-unitary dynamics through state-vector simulations, the concrete implementation of the non-unitary operator $(I - \Delta\tau H_p)$ on actual quantum hardware constitutes a significant open challenge. Since physical quantum processors are fundamentally restricted to unitary gate operations, executing this non-unitary step requires alternative strategies, such as utilizing ancilla qubits to construct a globally unitary evolution that projects the desired non-unitary action locally onto the system, a mechanism closely analogous to block encoding. However, a major bottleneck arises in its sequential application across multiple algorithmic layers: the method relies on preparing the ancillas in a superposition state followed by projective measurements and post-selection at each iteration, causing the cumulative success probability to decay drastically over successive steps. Consequently, our ongoing and future efforts are focused on optimizing the quantum circuit compilation to mitigate or circumvent the requirement for continuous sequential measurements, or alternatively, exploring variational quantum imaginary time evolution (QITE) frameworks. Successfully resolving these compilation bottlenecks remains essential for transitioning the ITE-TR-FALQON protocol from an algorithmic proof-of-principle into a viable, hardware-level simulation framework.

In summary, this study evidences that the combination of deterministic control strategies (Lyapunov) with controlled dissipation (ITE) offers a promising path for quantum simulation. ITE-TR-FALQON presents itself as a robust alternative to VQE, eliminating the dependence on classical optimization loops and mitigating problems such as Barren Plateaus, paving the way for scalable investigations of quantum gravity and condensed matter on NISQ devices.

\begin{acknowledgments}
G. E. L. P. acknowledges support from the Fundação de Amparo à Pesquisa do Estado de São Paulo (FAPESP), grant no.~2025/22498-8 (fellowship associated with grant no.~2024/08433-8). 
L. A. M. R. acknowledges support from FAPESP, grant no.~2025/25024-7. 
F. F. F. acknowledges support from FAPESP, grant nos.~2024/00998-6 (CRISQuaM -- FAPESP Research, Innovation and Dissemination Center, RIDC/CEPID) and 2025/15490-0 (QuantaNet -- FAPESP Thematic Project Grant). F. F. F. also acknowledges partial financial support from the Conselho Nacional de Desenvolvimento Científico e Tecnológico (CNPq) through the National Institute of Science and Technology for Applied Quantum Computing, Grant No.~408884/2024-0, and from the Office of Naval Research (ONR), Project No.~N62909-24-1-2012.
D.~R. acknowledges FAPESP for the ICTP-SAIFR grant 2021/14335-0 and the Young Investigator grant 2023/11832-9.
D.~R. also acknowledges the Simons Foundation for the Targeted Grant to ICTP-SAIFR.
A. R. F. acknowledges support from FAPESP, grant no.~2024/21658-9 (fellowship associated with grant no.~2024/08433-8). 
\end{acknowledgments}

\bibliography{main}

\end{document}